\documentclass[english,prb,preprint]{revtex4-1} 

\usepackage{amsmath}  
\usepackage{amsfonts} 
\usepackage{graphicx} 

\begin{document}

\title{Incompleteness of the Hamilton-Jacobi theory}
\author{Nivaldo A. Lemos\\
{\small{Departamento de F\'{\i}sica}}\\
{\small{Universidade Federal Fluminense}}\\
{\small{Av. Litor\^anea s/n, Boa Viagem - CEP 24210-340}}\\
{\small{Niter\'oi - Rio de Janeiro}}\\
{\small{Brazil}}\\
E-mail: nivaldo@if.uff.br }

\begin{abstract}

The problem of the motion of a charged  particle in an electric dipole field is used to illustrate  that the Hamilton-Jacobi method does not necessarily give all solutions to the equations of motion of a mechanical system. The mathematical reason for this phenomenon is discussed. In the particular case under consideration, it  is shown how to circumvent the difficulty and find the missing solutions by means of a very special limiting procedure. 

\end{abstract}

\maketitle

\section{Introduction}
 The Hamilton-Jacobi theory is the pinnacle of classical mechanics. It provides the most powerful method of integration of the equations of motion by reducing the task of solving $2n$ first-order ordinary differential equations (Hamilton's equations) to that of finding a certain particular solution (a complete integral)  to a single first-order partial differential equation in $n+1$ variables (the Hamilton-Jacobi equation). The deepest structural problems of classical mechanics are best investigated by the Hamilton-Jacobi theory. Perturbation theory and the celebrated KAM theorem, as well as action-angle variables and the theory of adiabatic invariants, are formulated in  terms  of the Hamilton-Jacobi formalism.\cite{Jose} Integrable systems, which never present chaotic behavior, are characterized as those systems whose equations of motion are solved by quadratures by the Hamilton-Jacobi method. 
 
 The standard texbooks on classical mechanics do not seem to call attention to the possibility that  the Hamilton-Jacobi theory may be  incomplete, in the sense  that  a complete integral  obtained by separation of variables may fail to produce some  solutions of Hamilton's equations, although this may have long been known by some. In the present work the problem of  the motion of a charged  particle in an electric dipole field is studied as a case in which  the Hamilton-Jacobi theory is actually incomplete, that is, there are  nontrivial solutions to Hamilton's equations that are beyond the reach of the standard Hamilton-Jacobi method based on the construction of a complete integral by separation of variables in the form of a sum. The mathematical reason for this phenomenon is identified. The problem of a charged  particle in the field of a point electric dipole is chosen not because it is particularly significant in itself, but mainly  because the relevant calculations can be performed explicitly and quite simply, allowing the clear identification of the reason why the Hamilton-Jacobi method fails and  suggesting how to find the missing solutions. In fact, for the example investigated a not wholly satisfactory procedure is proposed to find the missing solutions without entirely leaving  the realm of the Hamilton-Jacobi theory.

\section{Summary of the Hamilton-Jacobi theory}

Consider a mechanical system with $n$ degrees of freedom described by the canonical variables $(q,p)$ and  Hamiltonian $H(q,p,t)$. The aim of the Hamilton-Jacobi theory is the construction of a canonical transformation such that the new canonical variables $(Q,P)$ are constants of the motion.\cite{Landau, Goldstein} This is achieved by requiring that the transformed Hamiltonian $K(Q,P,t)$ be zero.
Assuming the generating function $S$ of the intended canonical transformation is a function of the old coordinates and the new momenta (a generating function of the type $F_2$ according to standard notation), $K=0$ is accomplished if  $S$ satisfies the Hamilton-Jacobi equation
\begin{equation}
\label{HamiltonJacobi}
H\Bigl( q_1, \ldots ,q_n, \frac{\partial S}{\partial q_1}, \ldots , \frac{\partial S}{\partial q_n}, t\Bigr) + \frac{\partial S}{\partial t} =0 \, .
\end{equation}  

For the purpose of solving Hamilton's equations for the original canonical variables the general solution to Eq. (\ref{HamiltonJacobi}) is not needed. All one needs is a {\it complete integral}, that is, a particular solution $S(q_1, \ldots ,q_n, \alpha_1, \ldots , \alpha_n, t)$ containing $n$ independent and non-additive parameters, which are identified with the new momenta (this identification is consistent because, by construction, the new momenta are constants of the motion). The precise meaning of the previous qualifications on the parameters $\alpha_1, \ldots , \alpha_n$ is spelled out by the following definition.

\bigskip

{\bf Definition.} A {\it complete integral} of the Hamilton-Jacobi equation   (\ref{HamiltonJacobi}) is a particular solution $S(q_1, \ldots ,q_n, \alpha_1, \ldots , \alpha_n, t)$ containing $n$ arbitrary constants $\alpha_1,\ldots , \alpha_n$ and such that
\begin{equation}
\label{Nonvanishing-Determinant}
\det \bigg( \frac{\partial^2 S }{\partial \alpha_i \partial q_j}\bigg) \neq 0 \, .
\end{equation}   

\bigskip

Once in possession of a complete integral $S(q, \alpha ,t)$ to the Hamilton-Jacobi equation, one solves Hamilton's equations for $(q,p)$   by means of
\begin{equation}
\label{beta-i}
\beta_i = \frac{\partial S(q, \alpha ,t)}{\partial \alpha_i}
\end{equation}   
where $\beta_1, \ldots , \beta_n$ are constants, and
\begin{equation}
\label{p-i}
p_i = \frac{\partial S(q, \alpha ,t)}{\partial q_i} \, .
\end{equation} 
Jacobi's theorem states  that this method always furnishes solutions to  Hamilton's equations.

\bigskip

{\bf Theorem (Jacobi).} Let $S(q, \alpha ,t)$ be a complete integral to the Hamilton-Jacobi equation (\ref{HamiltonJacobi}). If $\beta_1, \ldots, \beta_n$ are constants, then $q_i(t)$ and $p_i(t)$ determined by Eqs. (\ref{beta-i}) and (\ref{p-i}) satisfy Hamilton's equations

\begin{equation}
\label{Hamiltons-equations}
{\dot q_i} = \frac{\partial H}{\partial p_i} \, , \,\,\,\,\,\,\,\,\,\, {\dot p_i} = -\frac{\partial H}{\partial q_i}\,  .
\end{equation} 

\bigskip
 
 A proof of this theorem can be found, for example,  in Refs. 4 and 5. 
 
 According to the implicit function theorem, condition (\ref{Nonvanishing-Determinant}) insures that the $n$ Eqs. 
(\ref{beta-i}) can be solved for the coordinates in the form $q_i(t) = f_i(\alpha, \beta, t)$. Insertion of these functions into the right-hand side of Eq. (\ref{p-i}) yields $p_i(t) = g_i (\alpha, \beta, t)$. Thus one finds the coordinates and momenta as functions of time and $2n$ arbitrary constants, namely the $n$ alphas and the $n$ betas. Since the initial conditions are also $2n$ in number, to wit the values of the $n$ coordinates  and $n$  momenta at some initial time $t=t_0$, it is taken for granted that the alphas and  betas can always be chosen to accommodate all initial conditions whatsoever, so that the Hamilton-Jacobi theory gives all solutions to Hamilton's equations of motion. But in order for this to be the case condition (\ref{Nonvanishing-Determinant}) is crucial, as we presently explain.

With $t=t_0$ Eq. (\ref{p-i}) becomes

\begin{equation}
\label{p-i-zero}
p_{0i} = \frac{\partial S}{\partial q_i}(q_0, \alpha ,t_0) \, ,
\end{equation} 
where $(q_0,p_0)$ are the initial values of $(q,p)$. Condition (\ref{Nonvanishing-Determinant}) guarantees that these equations can be solved for $\alpha_1, \ldots , \alpha_n$ in terms of $q_0, p_0, t_0$. Insertion of the alphas so obtained into the right-hand side of   Eq. (\ref{beta-i}) determines $\beta_1, \ldots , \beta_n$ in terms of $q_0, p_0, t_0$. Therefore, if condition (\ref{Nonvanishing-Determinant}) fails for certain values of $\alpha_1, \ldots ,\alpha_n, q_{01}\, \ldots , q_{0n}$, some initial conditions may be inaccessible whatever the choice of the alphas and betas, which  means that some solutions to the equations of motion are missed. It is worth stressing that condition (\ref{Nonvanishing-Determinant}) may fail  because either $\det (\partial^2 S/\partial \alpha_i \partial q_j)$ is zero or does not exist for some  values of $\alpha_1, \ldots , \alpha_n, q_1, \ldots , q_n$.

 We proceed to show, by means of a physically interesting example, that the standard Hamilton-Jacobi method, based on the construction of a complete integral to the Hamilton-Jacobi equation by separation of variables, does not always give the totality of  solutions to the equations of motion.

\section{Charged particle in the field of an electric dipole} 

Let us examine the motion of a charged particle in the electric field created by a point electric dipole fixed at the origin. Since Lagrange's and Hamilton's equations are equivalent, we first discuss the equations of motion from the Lagrangian point of view. Next we study the solution of  the equations of motion by the Hamilton-Jacobi method.

\subsection{Lagrangian approach}

The electrostatic potential energy of a particle with electric charge $q$ in the field of a point electric dipole of magnitude $p_0$ fixed at the origin and oriented in the $z$-direction is $V(r, \theta)= k\cos \theta/r^2$, where $\,r, \theta, \phi\,$  are spherical coordinates and $k=qp_0/4\pi\epsilon_0$. Therefore, the particle's motion is described by the  Lagrangian 

\begin{equation}
\label{Lagrangian}
L= T-V = \frac{m}{2} \bigl(  {\dot r}^2 + r^2{\dot \theta}^2 + r^2 \sin^2 \theta \,{\dot \phi}^2 \bigr) - \frac{k\cos \theta}{r^2}\,  .
\end{equation} 
Lagrange's equations are

\begin{equation}
\label{Lagrange-equation-r}
m{\ddot r} -mr {\dot \theta}^2 -  m r \sin^2 \theta \,{\dot \phi}^2 - \frac{2k\cos\theta}{r^3} =0\,  ,
\end{equation} 

\begin{equation}
\label{Lagrange-equation-theta}
mr^2{\ddot \theta} + 2m r {\dot r}{\dot \theta} -  m r^2  \sin\theta \cos\theta \,{\dot \phi}^2 - \frac{k\sin\theta}{r^2}  = 0\,  ,
\end{equation}

\begin{equation}
\label{Lagrange-equation-phi}
\frac{d}{dt} \bigl(m r^2 \sin^2 \theta \,{\dot \phi} \bigr) =0 \,\, \Longrightarrow \,\, m r^2 \sin^2 \theta \,{\dot \phi} = \ell = \mbox{constant} \, ,
\end{equation}
where $\ell$ denotes the constant value of $L_z$, the $z$-component of the particle's angular momentum with respect to the origin.

If $\ell =L_z=0$ there is a remarkable solution\cite{Jones} to these equations of motion, namely   $\phi =0$, $r=r_0$ with the angle $\theta$ satisfying
\begin{equation}
\label{Equation-theta-ddot}
mr_0^2 {\ddot \theta}  - \frac{k\sin\theta}{r_0^2}  = 0
\end{equation}
and 
\begin{equation}
\label{Equation-theta-dot}
mr_0 {\dot \theta}^2 =   -\frac{2k\cos\theta}{r_0^3} \, .
\end{equation}
For $\phi =0$ the range of values of $\theta$ has to be redefined to $[0,2\pi )$ in order that the coordinates $(r, \theta )$  cover the entire $xz$-plane. Assuming $k>0$, Eq. 
(\ref{Equation-theta-dot}) requires $ \frac{\pi}{2} \leq \theta \leq \frac{3\pi}{2}$. Putting $\psi = \theta - \pi$ we have  $ -\frac{\pi}{2} \leq \psi \leq \frac{\pi}{2}$ and Eq. (\ref{Equation-theta-ddot}) becomes
\begin{equation}
\label{Equation-psi-ddot}
{\ddot \psi}  + \frac{k}{mr_0^4}\sin\psi  = 0 \, ,
\end{equation}
which is the equation of a pendulum. From Eq. (\ref{Equation-theta-dot}) it follows that the turning points are $ \theta = \pi /2$ and  $ \theta = 3\pi /2$, which correspond to 
$ \psi = -\pi /2$ and  $ \psi = \pi /2$. The particle moves in a semicircular path  on the half-plane $z \leq 0$ exactly like a pendulum with length $r_0$ and amplitude $\pi/2$,  oscillating periodically between the point with coordinate $r_0$ and the point  with coordinate $-r_0$, both on   the $x$-axis. From Eq. (\ref{Equation-theta-dot})  it also follows that this solution  has zero energy:
\begin{equation}
\label{Energy}
E=T+V= \frac{mr_0^2}{2}{\dot \theta}^2  +  \frac{k\cos \theta}{r_0^2} = -\frac{k\cos\theta}{r_0^2} + \frac{k\cos \theta}{r_0^2} = 0 \, .
\end{equation}

Now we go on to establish that this solution cannot be obtained directly by the standard Hamilton-Jacobi formalism.

\subsection{Hamilton-Jacobi approach}

The Hamiltonian associated with  the Lagrangian (\ref{Lagrangian}) is 

\begin{equation}
\label{Hamiltonian}
H= \frac{p_r^2}{2m} + \frac{p_{\theta}^2}{2mr^2} + \frac{p_{\phi}^2}{2mr^2 \sin^2\theta}  + \frac{k\cos \theta}{r^2}\, ,
\end{equation} 
and the corresponding Hamilton-Jacobi equation is
\begin{equation}
\label{Hamiltonian-Jacbi-equation-Kepler}
\frac{1}{2m}\bigg( \frac{\partial S}{\partial r}\bigg)^2 + \frac{1}{2mr^2}\bigg( \frac{\partial S}{\partial \theta}\bigg)^2 + \frac{1}{2mr^2 \sin^2\theta}\bigg( \frac{\partial S}{\partial \phi}\bigg)^2 + \frac{k\cos \theta}{r^2} +  \frac{\partial S}{\partial t} = 0\, .
\end{equation} 
This equation can  be solved by separation of variables in the form\cite{Landau,Goldstein}
\begin{equation}
\label{Separation-HJ}
S= -Et + \alpha_{\phi} \phi + W_1(r) + W_2(\theta )\, ,
\end{equation}
where the separation constant $E$ is the total energy and the second separation constant $\alpha_{\phi}$ is the $z$-component of the angular momentum, previously denoted by $\ell$.  
Insertion of Eq. (\ref{Separation-HJ}) into Eq. (\ref{Hamiltonian-Jacbi-equation-Kepler}) followed by multiplication by $r^2$ leads to 
\begin{equation}
\label{Separated-W1-W2}
r^2\bigg[ \frac{1}{2m}\Bigl( \frac{dW_1}{dr}\Bigl)^2  -E\bigg]= - \bigg[ \frac{1}{2m}\Bigl( \frac{dW_2}{d\theta}\Bigl)^2 + \frac{\alpha_{\phi}^2}{2m\sin^2\theta} + k\cos \theta\bigg] = -\frac{\alpha_{\theta}}{2m}\, ,
\end{equation}
where the third separation constant is conveniently written as  $\alpha_{\theta}/2m$.
It is straightforward to solve the two ordinary differential equations (\ref{Separated-W1-W2}) for $W_1$ and $W_2$ to obtain the following complete integral to the Hamilton-Jacobi equation  (\ref{Hamiltonian-Jacbi-equation-Kepler}):

\begin{equation}
\label{Complete-integral}
S(r,\theta ,\phi ,E,\alpha_{\theta},\alpha_{\phi},t)= -Et + \alpha_{\phi} \phi + \int \Bigl[ 2m E - \frac{\alpha_{\theta}}{r^2}\Bigr]^{1/2} dr   + \int \Bigl[\alpha_{\theta} - 2mk\cos\theta  - \frac{\alpha_{\phi}^2}{\sin^2 \theta}\Bigr]^{1/2} d\theta \, .
\end{equation}

This complete integral solves the equations of motion by quadratures:

\begin{equation}
\label{beta-one}
\beta_1 = \frac{\partial S}{\partial E} =  -t +  \int \frac{m dr}{ \Bigl[ 2m E- \alpha_{\theta}/r^2\Bigr]^{1/2}}\, ;
\end{equation}

\begin{equation}
\label{beta-two}
\beta_2 = \frac{\partial S}{\partial \alpha_{\theta}} =  - \frac{1}{2}\int \frac{dr}{r^2 \Bigl[ 2m E - \alpha_{\theta}/r^2\Bigr]^{1/2}} + \frac{1}{2} \int \frac{d\theta}{ \Bigl[\alpha_{\theta} -2mk\cos\theta - \alpha_{\phi}^2/\sin^2 \theta \Bigr]^{1/2}} \, ;
\end{equation}

\begin{equation}
\label{beta-three}
\beta_3 = \frac{\partial S}{\partial \alpha_{\phi}} = \phi  - 
\int \frac{\alpha_{\phi} d\theta}{\sin^2\theta \Bigl[\alpha_{\theta} -2mk\cos\theta - \alpha_{\phi}^2/\sin^2 \theta \Bigr]^{1/2}}\, .
\end{equation}

Equation  (\ref{beta-one}) seemingly allows us to completely determine the radial motion $r(t)$. There are three cases to consider.

Case (i): $E>0$. In this case we can write Eq. (\ref{beta-one}) in the form
\begin{equation}
\label{radial-E-positive}
\beta_1 =  -t +  \sqrt{\frac{m}{2E}}\int \frac{rdr}{\sqrt{r^2-a}}=\sqrt{\frac{m}{2E}}\, {\sqrt{r^2-a}}\, ,
\end{equation}
whence
\begin{equation}
\label{radial-E-positive-explicit}
r= \bigg[ a+ \frac{2E}{m} (t+ \beta_1)^2 \bigg]^{1/2}\, ,  \,\,\,\,\,\, a=\frac{\alpha_{\theta}}{2mE}\,  .
\end{equation}

Case (ii): $E=0$. This requires $\alpha_{\theta} < 0$ and  Eq. (\ref{beta-one}) leads to 
\begin{equation}
\label{radial-E-zero}
\beta_1 =  -t +  \frac{m}{ \vert\alpha_{\theta}\vert^{1/2}}\int rdr = -t +  \frac{m}{\vert\alpha_{\theta}\vert^{1/2}} \frac{r^2}{2}
\end{equation}
which implies
\begin{equation}
\label{radial-E-zero-explicit}
r= \sqrt{\frac{2{\vert\alpha_{\theta}\vert^{1/2}}}{m} }(t+ \beta_1)^{1/2}\, .
\end{equation}

Case (iii): $E<0$. This also requires $\alpha_{\theta} < 0$. Then equation  Eq. (\ref{beta-one}) becomes
\begin{equation}
\label{radial-E-negative}
\beta_1 =  -t +  \sqrt{\frac{m}{2\vert E\vert }}\int \frac{rdr}{\sqrt{a -r^2}}= -t - \sqrt{\frac{m}{2\vert E\vert}}{\sqrt{a - r^2}}\, ,
\end{equation}
and it follows that
\begin{equation}
\label{radial-E-negative-explicit}
r= \bigg[ a - \frac{2\vert E\vert }{m} (t+ \beta_1)^2 \bigg]^{1/2}\, , \,\,\,\,\,\, a=\frac{\vert \alpha_{\theta}\vert}{2m\vert E\vert }\,  .
\end{equation}

Note that in no case does one find a constant $r$.  If  attacked  by the Hamilton-Jacobi method alone, the problem of the motion of a charged  particle in an electric dipole field does not reveal
the beautiful solution found by Jones.\cite{Jones}

The trouble arises from the  circumstance that the pendulum-like motion takes place with  $E=0$, $\alpha_{\phi}=0$ and, because of Eq. (\ref{Separated-W1-W2}),  $\alpha_{\theta}=0$, since $dW_1/dr=p_r=m{\dot r}=0$ for $r=r_0$.  For these values of the separation constants Eq. (\ref{beta-three}) gives the correct result $\phi =\beta_3 = \mbox{constant}$, but Eqs. (\ref{beta-one})
and (\ref{beta-two}) are not defined.

Another difficulty can be pointed out.  Note that Eq. (\ref{beta-one}) furnishes directly time as a function of the radial coordinate, namely $t(r)$, the inverse function of $r(t)$. Since a constant function has no inverse,  Eq. (\ref{beta-one}) cannot give the solution $r=r_0$ shown to exist by the Lagrangian formalism. Definitely, the pendulum-like solution cannot be obtained by the standard
Hamilton-Jacobi method.

\section{What went wrong?}

Differentiating Eq. (\ref{beta-one}) with respect to time we get

\begin{equation}
\label{time-derivative-beta-one}
1=  \frac{m{\dot r}}{\Bigl[ 2m E - \alpha_{\theta}/r^2\Bigr]^{1/2}} \, .
\end{equation}
Formally, $r=\mbox{constant}$ may be compatible with the above equation only if
\begin{equation}
\label{compatibility-r-constant}
2m E - \frac{\alpha_{\theta}}{r^2} = 0 \, .
\end{equation} 


Let us take a look at what this condition   implies about the complete integral (\ref{Complete-integral}). 
Setting $q_1=r, q_2=\theta, q_3=\phi , \alpha_1 =E, \alpha_2 =\alpha_{\theta},\alpha_3 =\alpha_{\phi}  $, it follows immediately from Eqs. (\ref{beta-one})-(\ref{beta-three}) that 
\begin{equation}
\label{matrix-complete-integral}
\bigg( \frac{\partial^2 S}{\partial \alpha_i \partial q_j} \bigg) = 
 \left( \begin{array}{ccc}
       \frac{m}{ \bigl( 2m E - \alpha_{\theta}/r^2\bigr)^{1/2}} & 0 & 0 \\ \\
       - \frac{1}{2r^2 \bigl( 2m E - \alpha_{\theta}/r^2\bigr)^{1/2}} &      \frac{1}{2\bigl(\alpha_{\theta} - 2mk\cos\theta  - \alpha_{\phi}^2/\sin^2 \theta \bigr)^{1/2}} & 0 \\ \\
       0 & -\frac{\alpha_{\phi}}{\sin^{2}\theta \bigl(\alpha_{\theta} - 2mk\cos\theta  - \alpha_{\phi}^2/\sin^2 \theta \bigr)^{1/2}} & 1
       \end{array}\right) \, ,
\end{equation}
from which one finds at once
\begin{equation}
\label{determinant-matrix-complete-integral}
\det \bigg( \frac{\partial^2 S}{\partial \alpha_i \partial q_j} \bigg) = (m/2) 
\Bigl[ 2m E - \alpha_{\theta}/r^2\Bigr]^{-1/2} \, 
       \Bigl[\alpha_{\theta} - 2mk\cos\theta - \alpha_{\phi}^2/\sin^2 \theta \Bigr]^{-1/2} \, .
\end{equation}
This determinant is not defined (it is formally infinite) whenever  condition  (\ref{compatibility-r-constant})  holds and, as a consequence, $S$ given by  Eq. (\ref{Complete-integral}) is not a bona fide complete integral to the Hamilton-Jacobi equation. The same problem occurs if $E=0$ and $\alpha_{\theta}=0$ whatever the values of $\alpha_{\phi}, r, \theta, \phi$. The pendulum-like motion   corresponds to a situation in which the particular solution to the Hamilton-Jacobi equation obtained by separation of variables fails to be a complete integral. 

\section{Finding the missing solution}

Although, strictly speaking, the pendulum-like motion with $r=r_0$ and  $\phi = 0$ cannot be obtained directly from Eqs. (\ref{beta-one}) to (\ref{beta-three}), there is an indirect way to get this motion.

Taking the time derivative of Eq. (\ref{beta-two}) we find

\begin{equation}
\label{time-derivative-beta-two}
\frac{\dot\theta}{ \Bigl[\alpha_{\theta} -2mk\cos\theta  - \alpha_{\phi}^2/\sin^2 \theta \Bigr]^{1/2}} = \frac{\dot r}{r^2 \Bigl[ 2m E - \alpha_{\theta}/r^2\Bigr]^{1/2}}  \, .
\end{equation}
Assuming that Eq. (\ref{compatibility-r-constant}) does not hold, division of Eq. (\ref{time-derivative-beta-two}) by Eq. (\ref{time-derivative-beta-one}) gives
\begin{equation}
\label{division-time-derivative-beta-two-one}
\frac{\dot\theta}{ \Bigl[\alpha_{\theta} - 2mk\cos\theta - \alpha_{\phi}^2/\sin^2 \theta \Bigr]^{1/2}} = \frac{1}{mr^2}  \, .
\end{equation}
Setting now $\alpha_{\phi}=0$ and taking the limit $E\to 0$ and $\alpha_{\theta} \to 0$ with the stipulation that the ratio  $\alpha_{\theta}/E$ remain fixed and positive, which gives $r=r_0$ either from Eq.  (\ref{radial-E-positive-explicit}) or Eq. (\ref{radial-E-negative-explicit}), we find
\begin{equation}
\label{time-derivative-theta}
\frac{\dot\theta}{ \sqrt{ - 2k\cos\theta}} = \frac{1}{\sqrt{m}\, r_0^2}  \, ,
\end{equation}
which is equivalent to  Eq. (\ref{Equation-theta-dot}).

Although successful, the above reasoning is highly unsatisfactory. By differentiating with respect to time  we were forced to undo what the Hamilton-Jacobi method had so remarkably achieved: the solution to the equations of motion in terms of integrals of known functions. In order to get $r=r_0$ we had to take the simultaneous limit $E \to 0$ and $\alpha_{\theta} \to 0$  in Eq. (\ref{radial-E-positive-explicit}) or Eq. (\ref{radial-E-negative-explicit}) with the very particular condition that the limit of the  ratio  $ \alpha_{\theta}/E$  be finite and positive. But taking this limit is suggested only by our previous knowledge that a constant  solution $r=r_0$ does exist.

\section{Conclusion}

Even when the Hamilton-Jacobi method succeeds in reducing the solution of Hamilton's equations to quadratures, its full success depends on the inversion of certain functions. If at least one inversion is impossible the method runs the risk of failing.  The example of  motion of a charged  particle in an electric dipole field suggests that impossibility of inversion may be a symptom that a solution to the equations of motion exists that cannot be directly reached by the method. Another such symptom, probably not independent of the former,  is the nonexistence of $\det (\partial^2 S/\partial \alpha_i \partial q_j)$ for certain values of $\alpha_1, \ldots , \alpha_n, q_1, \ldots , q_n$.

The fact, pointed out on the last paragraph of Section III, that a constant function has no inverse implies that the circular orbits for the classic Kepler problem are also missed by the standard Hamilton-Jacobi method. Some sort of limiting procedure is required to get these orbits, but it is not so straightforward to carry out explicitly because the integral in the equation corresponding to Eq. (\ref{beta-one}) leads to very complicated functional relations (see Section 3-8 of Goldstein\cite{Goldstein}). It is also worth mentioning that the problem discussed in the present paper admits other solutions\cite{Gutierrez} with $r=\mbox{constant}$ but nonvanishing angular momentum ($\alpha_{\phi}\neq 0$). These motions are also  beyond the reach of the standard Hamilton-Jacobi method.

Of course, the particular solution (\ref{Complete-integral}) to the Hamilton-Jacobi equation (\ref{Hamiltonian-Jacbi-equation-Kepler}) is not expected to be unique, most likely  there  are other solutions containing $n$ non-additive parameters. Perhaps there exists a complete integral of (\ref{Hamiltonian-Jacbi-equation-Kepler}) capable of generating the solutions that the complete integral (\ref{Complete-integral}) cannot. To our knowledge, other than separation of variables there is no systematic procedure to construct a complete integral to the Hamilton-Jacobi equation. However, once an $n$-parameter solution to the Hamilton-Jacobi  equation is found by separation of variables,  there is a somewhat inconspicuous scheme that allows the construction of new   $n$-parameter solutions.\cite{Epstein} If the solution found by separation of variables fails to satisfy condition (\ref{Nonvanishing-Determinant}) for some values of the separation  constants, that condition may happen to be satisfied for all values of the parameters by one of the new solutions.  For the problem studied here, it might be interesting to investigate whether the technique described by Epstein\cite{Epstein}  is able to give rise to a complete integral that generates the orbits   missed by the  
solution obtained by  separation of variables. 

In the case of motion of a charged  particle in an electric dipole field,  we showed how to find the missing solutions with some  violence to the spirit of the Hamilton-Jacobi theory, but it is not clear that the reasoning employed in this particular case will also work in other similar but more intrincate circumstances.

\end{document}